# Transforming unstructured voice and text data into insight for paramedic emergency service using recurrent and convolutional neural networks

Kyongsik Yun*, Thomas Lu, Alexander Huyen
Jet Propulsion Laboratory, California Institute of Technology
4800 Oak Grove Drive, Pasadena, CA 91109

**Abstract**

Paramedics often have to make lifesaving decisions within a limited time in an ambulance. They sometimes ask the doctor for additional medical instructions, during which valuable time passes for the patient. This study aims to automatically fuse voice and text data to provide tailored situational awareness information to paramedics. To train and test speech recognition models, we built a bidirectional deep recurrent neural network (long short-term memory (LSTM)). Then we used convolutional neural networks on top of custom-trained word vectors for sentence-level classification tasks. Each sentence is automatically categorized into four classes, including patient status, medical history, treatment plan, and medication reminder. Subsequently, incident reports were automatically generated to extract keywords and assist paramedics and physicians in making decisions. The proposed system found that it could provide timely medication notifications based on unstructured voice and text data, which was not possible in paramedic emergencies at present. In addition, the automatic incident report generation provided by the proposed system improves the routine but error-prone tasks of paramedics and doctors, helping them focus on patient care.

## 1. INTRODUCTION

Paramedics must make numerous life-saving decisions, often in the back of an ambulance with limited time. They sometimes ask the doctor for additional medical instructions, during which the patient's precious time is wasted. The Department of Homeland Security (DHS) Science and Technology Directorate (S&T) partnered with Canada's Department of National Defence Science and Technology Organization, Defence Research and Development Canada Centre for Security Science (DRDC CSS) to support this experiment to examine whether artificial intelligence could be used to improve the information overload. The experiment was part of the Next Generation First Responder Program by DHS [1–5].

Transcribing voice communications in paramedic emergency response environments is the first step for providing faster, more accurate first response to emergency patients. However, automatic speech recognition in this environment is particularly challenging due to the lack of training data, unfamiliar medical terms in acronyms, and noisy environment in emergency vehicles. We used bidirectional deep recurrent neural networks to train and test speech recognition performance. We showed that data augmentation and custom language models can improve speech recognition accuracy [6]. Then we used keywords extraction techniques to analyze spoken sentences and to summarize the next steps for treatment and for communication with doctors for the medical orders. The proposed system will help the machine analyze medical information in paramedic services and accelerate emergency medical responses.

Automatic speech recognition has three main applications, including input/output devices, communication aids, and information retrieval. The performance of automatic speech recognition has been greatly improved by using deep neural networks mainly for input / output devices [7]. Industry leaders such as Amazon, Apple, Baidu, Google, Microsoft, and IBM are evolving to help the public easily use their products through automatic speech recognition. Conversational interfaces based on Amazon's Alexa and Google's Home have been enhanced to naturally acquire information, access web services, and issue commands [8]. A recent study showed that deep learning based automatic speech recognition achieved human level performance in conversational speech recognition [9]. However, the accuracy of speech recognition in special purpose communication in noisy environments is not optimal [10].

*kyun@jpl.nasa.gov; phone 1 818 354-1468; fax 1 818 393-6752; jpl.nasa.gov

The language that paramedics use in the emergency vehicles is a unique language for patient care. They use many special terms and abbreviations that are difficult to decipher without prior knowledge of communication. In such an environment, speech recognition is important in terms of reducing the risk of misunderstanding, disseminating knowledge, and improving the patient care process. Therefore, general-purpose speech recognition may not work. To improve the accuracy of automatic speech recognition in the verbal communication of professionals, we need to consider a customized language model that can describe the relationship between words and abbreviations in sentences [11].

We believe that the combination of training data, deep neural networks, data augmentation, and a customized language model is the key to improving automatic speech recognition in the verbal communication of professionals.

In this study, we used a combination of technologies including speech recognition, sentence classification, lemmatization, parsing, keywords extraction, and optical character recognition to improve paramedic emergency response (Table 1). We consistently achieved the technology readiness levels of 4 and 5.

Table 1. Technology development summary.

| | | Accuracy | Speed (CPU macbook pro) | TRL |
|---|---|---|---|---|
| Main Technology | speech recognition (online) | ~0.9 | <1sec | 5 |
| | speech recognition (offline) | ~0.8 | 2-5sec | 4 |
| | sentence classification | ~0.8 | 2-5sec | 4 |
| | lemmatization | ~0.8 | <0.1sec | 4 |
| | parsing | ~0.8 | <0.1sec | 4 |
| | value extraction | ~0.8 | <0.1sec | 4 |
| | optical character recognition (online) | ~0.95 | <1sec | 5 |
| | optical character recognition (offline) | ~0.8 | <1sec | 4 |
| Method | Tesseract, Deepspeech, NLTK | | | |
| User | Paramedic | | | |
| Short term improvement | Lemmatization, parsing, value extraction, speech recognition (offline) | | | |
| Long term improvement | Background listening, streaming, real-time processing, wearable | | | |

## 2. TRAINING DATA AUGMENTATION

The recent development of deep learning-based speech recognition showed that conventional phoneme-based complex speech recognition is not required [12]. Rather, the system can output the final transcript directly based on the graphemes (i.e., alphabets) [13]. Previous studies used open-source data sets that typically contained thousands of hours of voice data [14]. State-of-the-art commercial speech recognition utilized 100,000 hours of data for training [15]. However, we have particularly small data sets because of the high cost of manual transcription (creation of ground truths) in specialized languages.

We trained our bidirectional recurrent neural networks using 76 unique sentences from paramedic emergency response. Nine lab members then created the voice data manually by saying the same sentence. We finally created a total ~2,000 speeches to analyze. Data augmentation techniques, including Gaussian noise (40 ~ 50dB), speech speed (0.9-1.1X), and volume modulation (-10 ~ 10dB) were applied to obtain augmented data corresponding to 10 times of original data [15,16]. Of the augmented data, the data including Gaussian noise was 60% and the speed and volume modulation data was 20% each.

## 3. BIDIRECTIONAL RECURRENT NEURAL NETWORKS

We used bidirectional recurrent neural networks (RNN) to train and test speech recognition performance. RNN can be thought of as an enhanced version of the hidden markov model (HMM) [13]. The HMM has a major drawback: the state is updated only from one state to the next so that the network cannot learn long-term dependencies [17]. For context-sensitive language decoding, it is important to learn long-term relationships between words. The solution is a recurrent neural network. Long-term dependence can be learned through back propagation [18].

Bidirectional RNNs collect information both in the past and in the future [19]. The original RNN, which obtains information entirely in the past, is not sufficient to accurately predict the current word, especially in situations where context is important. Given the sequence of words x(1), x(2), x(3), and x(4), the forward recurrent components can be represented by $a^{\rightarrow}(1)$, $a^{\rightarrow}(2)$, $a^{\rightarrow}(3)$, and $a^{\rightarrow}(4)$. x(1) is input to the forward recurrent component $a^{\rightarrow}(1)$. Output estimate is y(1). Then the backward layer of the recurrent components are $a^{\leftarrow}(4)$, $a^{\leftarrow}(3)$, $a^{\leftarrow}(2)$, and $a^{\leftarrow}(1)$. x(1) is also input to the backward recurrent component $a^{\leftarrow}(1)$ outputting y(1). The forward recurrent networks compute from $a^{\rightarrow}(1)$ to $a^{\rightarrow}(4)$ direction, and the backward recurrent networks compute from $a^{\leftarrow}(4)$ to $a^{\leftarrow}(1)$. The predicted output y(1) is the result of combined network components of $a^{\rightarrow}(1)$ and $a^{\leftarrow}(1)$. The entire network consists of acyclic graphs. For example, the output y(3) is based on the past x(1) and x(2), the current x(3), and the future x(4). We used 3 convolutional input layers, 6 recurrent layers, 1 fully connected layer, and 1 softmax layer [15,16].

We used long short-term memory (LSTM) as a recurrent network component. Conventional RNNs still have significant practical problems caused by the exponential decay of gradient descent that prevents learning long-term relationships between words. LSTM is a special type of recurrent neural network that can learn long-term dependencies through selective memory consolidation [20].

Then we used the Connectionist Temporal Classification (CTC) cost for speech recognition output [21]. With RNN, the input sequence matches one-to-one with the RNN output. Therefore, the number of outputs is large and redundant. In speech recognition, the number of input information is generally much larger than the number of output characters. For example, five seconds of a 16,000 Hz audio input is 80,000 inputs, whereas the number of output characters per 5 seconds is at least 1,000 times less than 80,000 characters. Therefore, the key is to reduce redundancy. The basic rule of CTC cost function is to erase repeated characters that are not separated by a "blank".

## 4. SEMANTIC LANGUAGE MODELING

We customized the language model to identify task-related abbreviations and technical terms. The language model uses the CTC output as input to return the probability of the last word in the given context of the previous words. The beam search algorithm was used to find the most accurate word candidates for speech recognition post-processing [22]. The algorithm considers three possible choices for the current word (beam width = 3). Unlike exact search algorithms, such as breadth first search or depth first search, beam search runs faster, but it is not guaranteed to find the exact maximum for $\arg\max_y P(y|x)$. Given the audio input x, y(t) is the output word at time t. Then the objective function is:

$$arg\ max_y \sum_{t=1}^{Ty} log\ P(y(t)\mid x,\ y(1),\ \dots,\ y(t-1))$$

The log was introduced to handle very small values of the probability, $P(y(t)\mid x,\ y(1),\ \dots,\ y(t-1))$, especially when the decoding sentence is long.

We found that bidirectional recurrent neural networks with data augmentation and custom language models achieved the best performance in a highly specialized language environment with specialized abbreviations and grammar structures. The result is significant in that even perfect human hearing perception cannot fully comprehend communication in the paramedic emergency response without prior knowledge. A vast amount of training and knowledge of specific situations is important for transcription of professional voice communications.

The speech recognition problem is to map the audio input x to the script y. The human ear converts a one-dimensional audio input to the intensity of the frequency component. This can be thought of as a preprocessing step to generate a spectrogram that maps 2D information of the time and frequency of the audio input. The human brain utilizes a variety of contextual information to fully understand and perceive speech, including attention [23] and social cues [24,25], as well as sound itself.

By using attention models in the future, we can bias the state of recurrent neural networks to improve context-based speech recognition by paying more attention to specific words and phrases [26]. The attention model must be trained as a separate RNN in the previous state to determine how much attention should be paid to adjacent inputs to bias the current state of the primary RNN. Naturally, the processing time to train the network will be at least doubled. Because of its inefficiency, the attention model should be carefully considered for speech recognition solutions. The attention model can also be used for machine translation [27] and image caption with visual attention [28].

Automatic speech recognition, supported by computer vision, can be useful for improving speech recognition accuracy in noisy environments. Previous studies have shown that lip reading computer vision is far superior to traditional noise reduction methods [29–31]. Another possibility that can improve accuracy is to use background information such as location information [32] and history information [33]. This information has been shown to improve speech recognition accuracy by reducing possible word combinations in certain situations.

## 5. PARAMEDIC EMERGENCY RESPONSE IMPROVEMENT

Challenges include understanding the conversation of paramedics and doctors, and translating spoken words into a patch form used by doctors when patients arrive at the hospital. Demonstrated capabilities include 1. publishing accurate and fast standing orders, 2. publishing accurately timed medication reminders, 3. no need to call a physician, 4. populating ePCR (electronic Patient Care Record) fields automatically and reducing time to generate reports. Ideal capabilities will be able to listen in the background, understand conversations, recognize speech from streaming audio, process and understand in real time, and provide a wearable platform and intuitive user interface (Table 2).

More specifically, a standing order was automatically generated based on the initial dispatch call, improving response time from an average of 3 minutes to 0 minutes. The paramedics then interact with the patient to identify the current medical condition, history, and potential treatment needed. However, the proposed model took longer than traditional emergency medical procedures. The main reason is that there was a problem with the user interface, and the actual paramedic was having trouble using the proposed model. This patient and paramedic interaction step is the most difficult part to improve using the proposed model. In the future, we believe that this can be improved through the development of an ideal model with features such as real-time speech processing, uninterrupted background continuous listening, and streaming audio analysis.

Medication dosage notification is a unique feature of the proposed model that was not possible during the existing emergency treatment process. Prescription errors are a serious problem that must be eliminated in the emergency response process of paramedics [34,35]. Paramedics are at risk of making mistakes, especially due to mental and physical overload on the job. The proposed model's reminder function helps to identify the right drug, the right dose, the right time and the right route. The benefits of this feature could not be quantitatively compared to traditional paramedic work given the limited sample size in this test.

In the proposed model, the medical condition, medical history, and treatment of the patient in an emergency was automatically shared with the doctor, so the paramedics did not have to call the doctor in the hospital for further medical orders. A previous study showed the possibility of using visualization techniques to improve communication and collaboration between paramedics and physicians [36]. In this study, doctors can verify patient information without a phone call with paramedics, and can provide medical orders immediately if needed. This automated process reduced call duration between doctors and paramedics from an average of 3 minutes to 0 minutes. However, generating a patch form and requesting additional medical orders took much longer in the proposed model than traditional emergency work. This was due to the inefficient user interface and lack of real-time processing capabilities. In ideal model estimation, we can improve the engineering part of the system, including the user interface and data processing efficiency, then spend almost zero time on patch form creation and medical order requests. The final stage of ePCR (electronic patient care record) generation, a post-incident report, improved from 10 minutes to 5 minutes. During the paramedic response, all actions and communication were recorded in real time in the proposed model, and a post-incident report is automatically generated, and the paramedic can simply review, revise and confirm the report.

Table 2. Performance comparisons among the existing system, proposed system (current), and estimated ideal system. Improvements are highlighted in yellow.

| Paramedic Emergency Response Steps | Existing Paramedic Work (min) | Proposed Model (min) | Estimated Ideal Model (min) | Notes |
|---|---|---|---|---|
| Dispatch | 0.75 | 0.75 | 0.75 | Same dispatch call process |
| Standing orders | 3 | 0 | 0 | Send standing orders without any delay |
| Paramedic arrival to incident | 10 | 10 | 10 | Physical travel time was the same. |
| Status | 1 | 5 | 1 | The proposed model took longer to enter data due to inconvenient user interface issues (real-time processing, background listening, streaming audio input capability required) |
| History | 2 | 7 | 2 | |
| Treatment | 2 | 10 | 2 | |
| Medication dosage reminder | none | Right drug, dosage, time, route | Right drug, dosage, time, route | A unique capability that was not possible in the current system |
| Paramedic call physician | 3 | 0 | 0 | Patient data was automatically shared with the doctor, so there was no need to call the doctor |
| Patch form | 5 | 7 | 0 | The proposed model took longer due to the inconvenient user interface for fixing errors in the patch form |
| Request to physician | 1 | 5 | 0 | |
| Physician order | 1 | 1 | 1 | The doctor reviewed the patch form and ordered treatment |
| ePCR (ACP) data input | 10 | 5 | 0 | The proposed model improved the post-incident report generation process |
| Total (min) | 38.75 | 50.75 | 16.75 | |

* ePCR: Electronic Patient Care Record
* ACP: Advanced Care Protocol

Finally, in the traditional paramedic work, the entire process took 38 minutes, 50 minutes from the proposed model, and 16 minutes from the expected ideal model. We believe that the estimated ideal model performance can be achieved after improving the user interface and real-time data processing capabilities of the proposed model in the future.

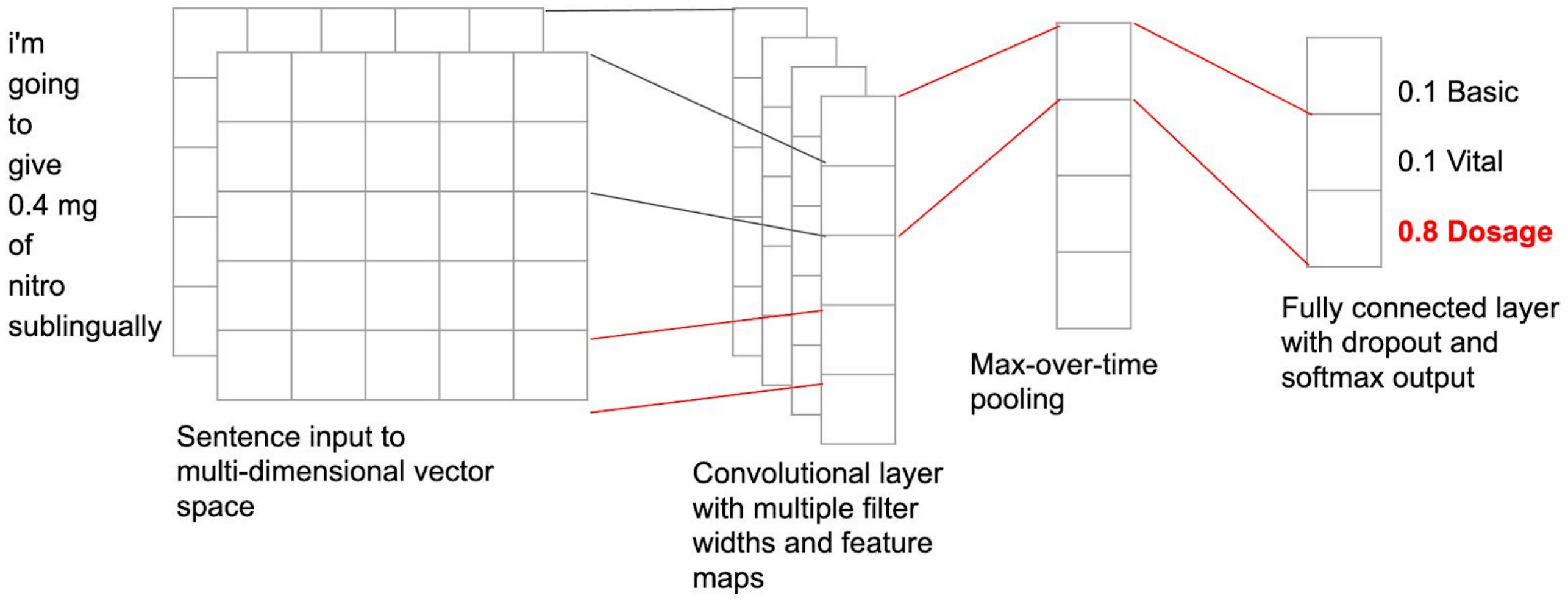

Figure 1. Convolutional neural networks to categorize sentences. The networks were trained on top of custom-trained word vectors for sentence-level classifications.

For speech understanding to support emergency response, converting speech to text was the first step, and to help further understand the sentence, each sentence had to be classified into three major classes. We organized three classes, including basic medical information, vital signs, and medication doses (Figure 1). The given sentence is transformed into a multidimensional vector matrix, and the input matrix goes through a convolutional layer with multiple filter widths and features, eventually sending three classes to the output [37].

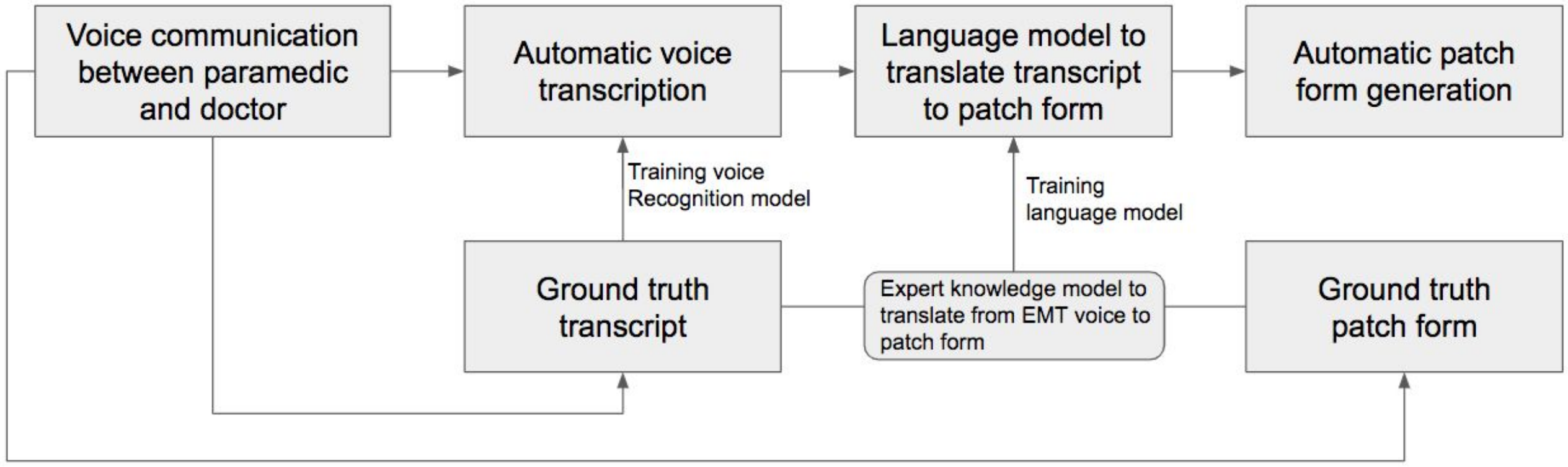

Figure 2. Complete patch form generation process 1. Generate a ground truth voice transcript through communication between paramedics and doctors. 2. Generate an expert knowledge model (doctor's mental logic) to translate from paramedic voice (ground truth transcript) to patch form. 3. Train the automatic voice transcription model. 4. Train the language model to translate the script into a patch using the expert model. 5. Generate an automatic patch form.

Lemmatization was implemented to improve text understanding using Python's NLTK (Natural Language Toolkit) [38]. The goal of lemmatization was to reduce inflectional forms and sometimes derivationally related forms of a word to a common base form. For instance, “am, are, is” are changed to “be”, and “car, cars, car's, cars'” to “car”. Using Python’s NLTK, we implemented a parser that can identify unnecessary words to remove them and key nouns to extract them. Patch form is the physician’s summary of the paramedic’s verbal report about the patient on scene. We implemented automatic generation of patch form information (audio summary) using paramedic speeches. The entire patch form generation process is illustrated in Figure 2.

A domain-specific language model that transforms voice scripts into patch forms used by emergency personnel and doctors was key to improving the accuracy of natural language understanding. Graph-based expert knowledge models were used to generate domain-specific language models. For example, if a patient is identified as having a heart problem, the relevant database is loaded and ECG (electrocardiogram) vital signs are analyzed. Depending on the ECG results, the likelihood of relevant keywords (e.g. sinus rhythm, sinus tach) appearing in paramedic and physician conversations increases, thus maximizing the accuracy of natural language understanding in the semantic language model. This process is very similar to the process of understanding human language in that it not only analyzes sounds of speeches, but also understands the context of a particular situation and the topic area of the mentioned subject.

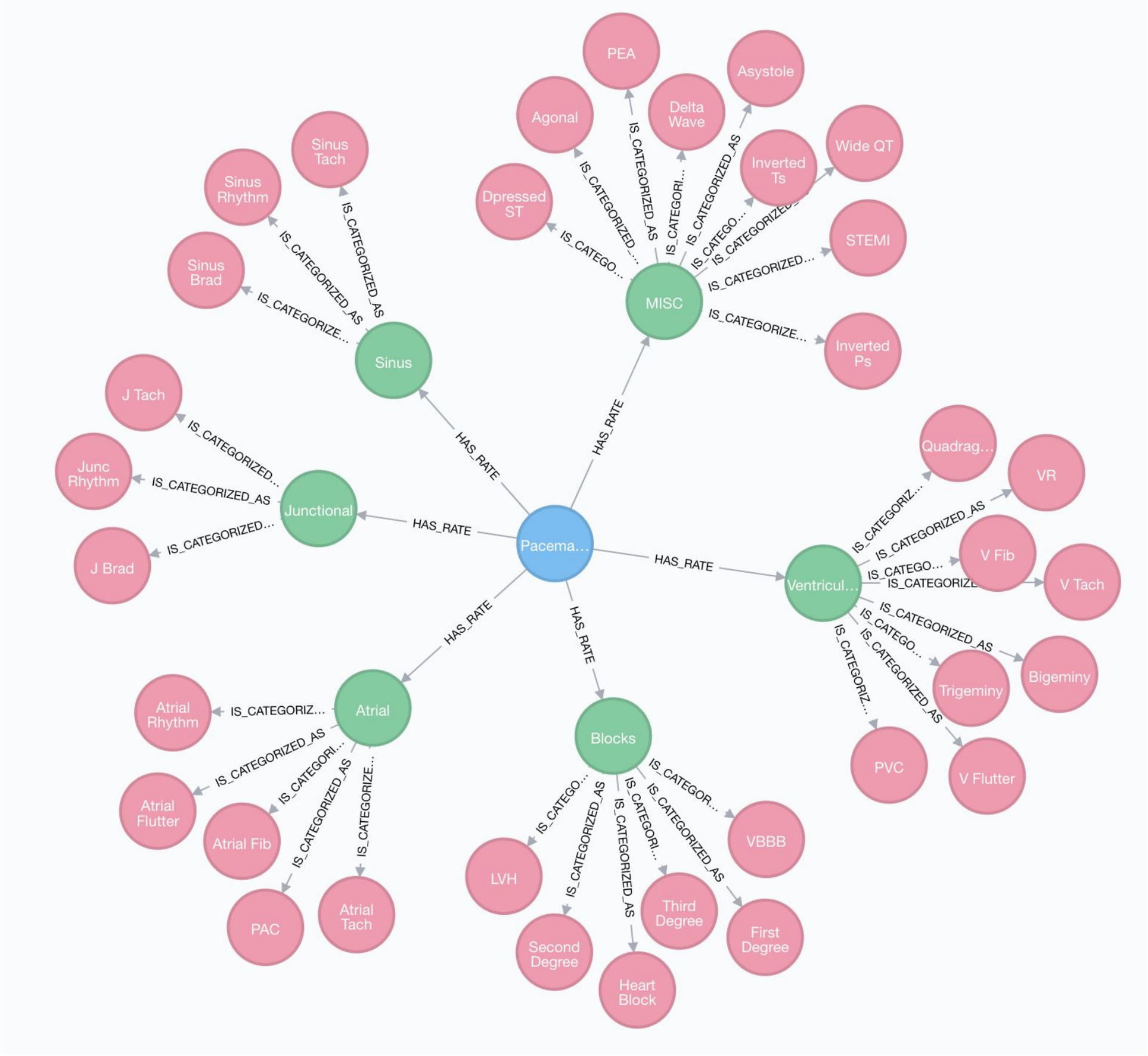

Figure 3. Graph database of terminology used in patching (paramedic-doctor communication). The database can be used to improve parsing and extracting information from unstructured text (speech transcript).

We achieved 93.3% test accuracy in speech recognition and understanding. Examples of voice transcription and keyword extraction are listed in Table 3 below. Keywords were extracted and categorized for further understanding, including previous nitroglycerin doses for treatment, abdominal physical findings, skin condition, past medical history, and medication reminders. In addition, basic medical information and vital signs were extracted for treatment recommendations and data transfer to the hospital. Basic medical information includes age, gender, Canadian Triage and Acuity Scale (CTAS), blood pressure, medication, pupil examination, temperature, pulse, allergies, physical examination and skin color.

Table 3. Representative examples of speech transcript and keywords extraction

| Paramedic_speech/ output | # correct | # all fields |
|---|---|---|
| { "transcript": ".requesting treatment of additional nitroglycerin", "treatment": "additional, nitroglycerin"} | 1 | 1 |
| {"transcript": ".abdomen is rigid abdomen is distended abdomen is percentile abdomen hasa scar running from the right side to the left side. ", "physical_findings_abdomen": "rigid, distended"} | 2 | 2 |
| {"transcript": ".allergies penicillin skin condition clammy history of mental illness ", "allergies": "Penicillin", "physical_findings_skin_condition": "clammy", "history": "mental_illness"} | 3 | 3 |
| { "transcript": ".paramedic request treatment of additional iv of saline", "treatment": "additional, IV"} | 1 | 1 |
| {"transcript": ". patient has history of nitroglycerin seizure psychiatrique and diabetes. ", "history": "nitroglycerin seizure psychiatrique and diabetes", "past_medical_history": "seizure, psychiatric, diabetes, nitroglycerin", "NTG_prior": "1"} | 3 | 3 |
| {"transcript": ".50 year old male patient complaining of substernal chest pain 0 shortness of brea the patient is at a single family residential address in belleville, ontario. . ctas assessment 1 findings 2 patient current medications rasa nitro slow k lasix . patient has no allergies. . physical exam finds pale sweaty shortness of brea tripod position . pulse is 90 strong. bp is 154 over 90 respirations. 90% temperature is 37. . pupil is 3 + reactive skin condition pale cool sweaty. ", "age": "50", "gender": "M", "CTAS": "1", "BP": "154 / 90", "systolic": "154", "diastolic": "90", "pain": "0", "medications": "A_S_A, furosemide", "medications_comment": "r slow k", "pupil_left": "3", "pupil_right": "3", "pupil_reactive_left": "1", "pupil_reactive_right": "1", "temperature": "37", "pulse": "90", "physical_exam": " pale sweaty shortness of brea tripod position", "allergies": "NKA", "physical_findings_skin_color": "pale", "pale": "1", "sweaty": "1"} | 15 | 16 |
| {"transcript": ".patient is a 29 year old male. . complaining of substernal chest pain . sweating . profusely patient seems to have a lot of chest pressure. . patient is . under cardiovascular medication 0 hasa history of diabetes 0 patient is taking insulin. . patients blood pressure is 120 over 80. . patience . blood sugar is . 174 ", "age": "29", "gender": "M", "BP": "120 / 80", "systolic": "120", "diastolic": "80", "medications": "A_S_A, insulin", "medications_comment": "h history diabetes", "history": "substernal_chest_pain", "past_medical_history": "cardiac, diabetes"} | 10 | 11 |
| **Total accuracy = 93.3%** | | |

## 6. FLOODED ROAD, MEDICINE BOTTLE, AND HAZARD DETECTION USING REAL TIME CONVOLUTIONAL NEURAL NETWORK CLASSIFICATION

We used publicly available CCTV camera feeds to identify hazardous situations on the road. We used 614 accessible public cameras in Ontario, Canada (https://511on.ca/). Two hundred and fifty CityCam datasets and 181 flood road images were used for training InceptionV3 neural network [39] (Figure 4) (video demo: https://youtu.be/H-HDkjbXpAs).

Tesseract Optical Character Recognition (OCR) using long short-term memory (LSTM) based modeling was used [40,41]. LSTM neural networks perform better than other alternative neural network architecture models for OCR, and outperform the classic character recognition algorithms used in traditional OCR. For example, the LSTM network achieved the best results in unsegmented connected handwriting recognition. The accuracy of the LSTM network is highly dependent on training data. In order to improve OCR accuracy, the image quality was intentionally degraded and used as training data. If Tesseract's LSTM recognizer fails to recognize a particular character sequence, it can make a decision with a regular static shape classifier. In our medicine bottle recognition application, we improved our keyword recognition from 65% to 93% using a custom language model and a DIN (Drug Identification Number) lookup (Figure 5).

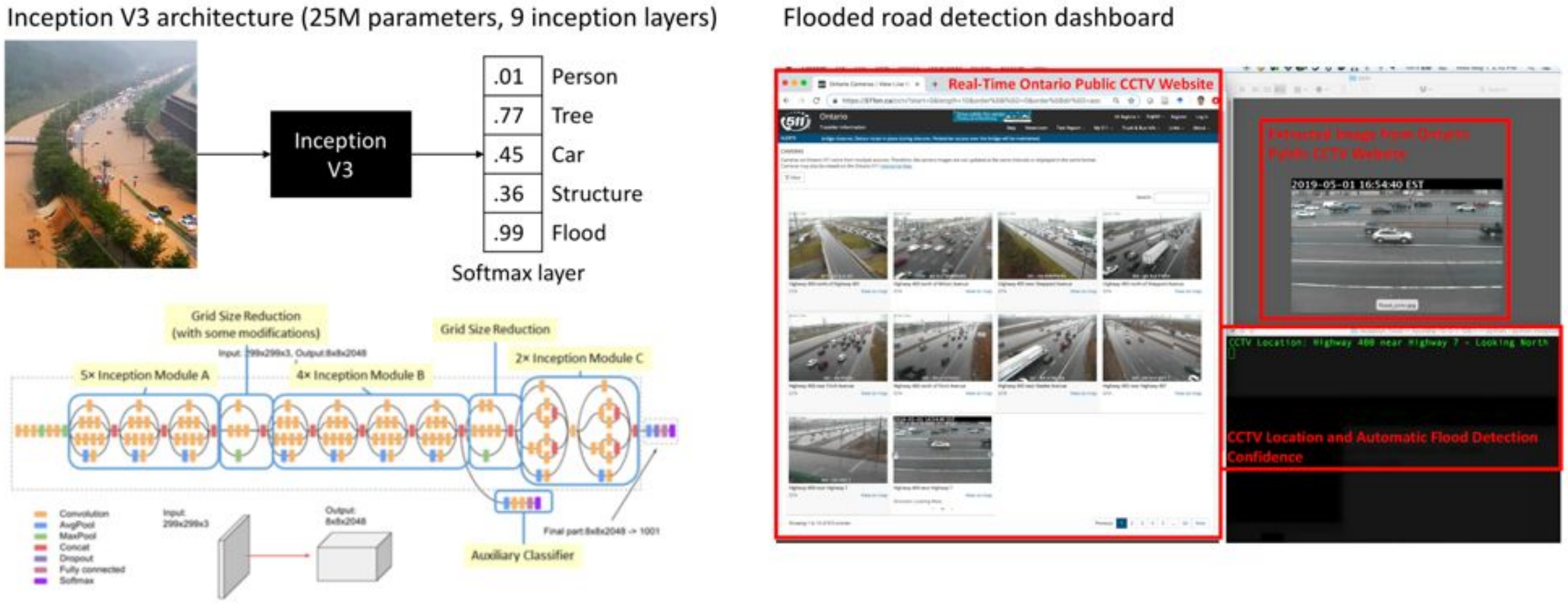

Figure 4. Inception V3 architecture for image classification, and flooded road detection dashboard showing real-time Ontario public CCTV website, extracted image from Ontario public CCTV website, and CCTV location and automatic flood detection confidence (video demo: https://youtu.be/H-HDkjbXpAs).

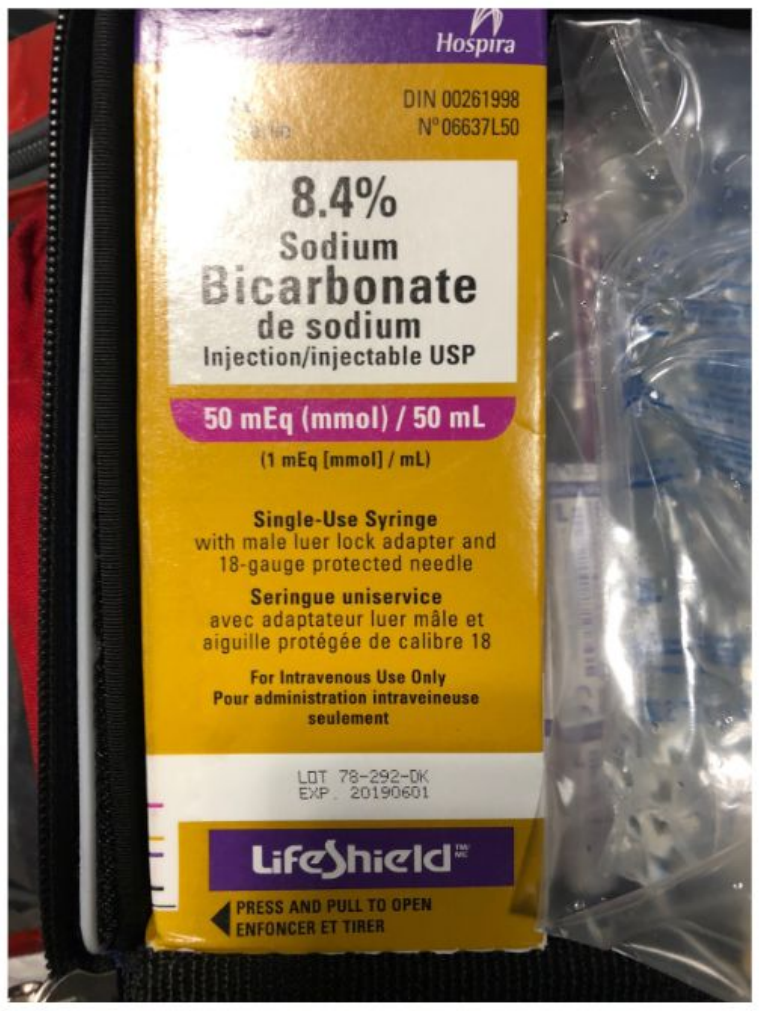

| Raw Transcript metrics: | CLM Transcript metrics: |
|---|---|
| Searching for keywords: ['DIN 00261998', '8.4%', 'Sodium', 'Bicarbonate', '50 mEq', '50 ml'] | Searching for keywords: ['DIN 00261998', '8.4%', 'Sodium', 'Bicarbonate', '50 mEq', '50 ml'] |
| **5/6 correct keywords: 83.33 %** **1 missed keywords: 16.67 %** | **6/6 correct keywords: 100.00 %** **0 missed keywords: 0.00 %** |
| Correctly identified keywords: ['DIN 00261998', '8.4%', 'Sodium', '50 mEq', '50 ml'] | Correctly identified keywords: ['DIN 00261998', '8.4%', 'Sodium', 'Bicarbonate', '50 mEq', '50 ml'] |
| Missed keywords: ['Bicarbonate'] | Missed keywords: [] |

Figure 5. Example of improvement in optical character recognition by custom language model (CLM)

Using the same OCR technology as used in medicine bottle recognition, we added a hazard placard recognition function to allow paramedics to quickly identify each hazard placard and send safety-related warnings (Figure 6). The hazard placard number identified in the Emergency Response Guidebook (ERG) is automatically entered to extract relevant information, and the information is then sent to paramedics.

We developed a real-time classifier that automatically retrieves and categorizes information from the database to assist in the rapid patient diagnosis of emergency personnel. The classifier was built on a historical record of generating recommendations for standardized medical orders using information such as patient age, patient gender, and patient's reported telephone calls. Paramedics refer to standing orders to determine treatment options according to the restrictions outlined in the document.

Paramedics' comments are automatically sent to the description field along with patient data collected by the initial dispatcher, and recommendations are generated when the minimum amount of information is met. The information collected through the call (between the dispatcher and patient) is stored and updated, and the standing order is automatically corrected if necessary.

Example input data: {'problem_nature_type': 'CHEST', 'problem_nature': 'Ischemic Chest Pain-(51)', 'gender': 'M', 'comment': '50YOM, SOB, pale diaphoretic, history of cardiac'}

Example output standing order: {'confidence_levels': [{'order': 'ACPE-ACP-2019', 'confidence': '0.086'}, {'order': 'CSMD-2019', 'confidence': '0.110'}, {'order': 'CIMD-ACP-2019', 'confidence': '0.804'}], 'timestamp': '20190101T010101-000000'}

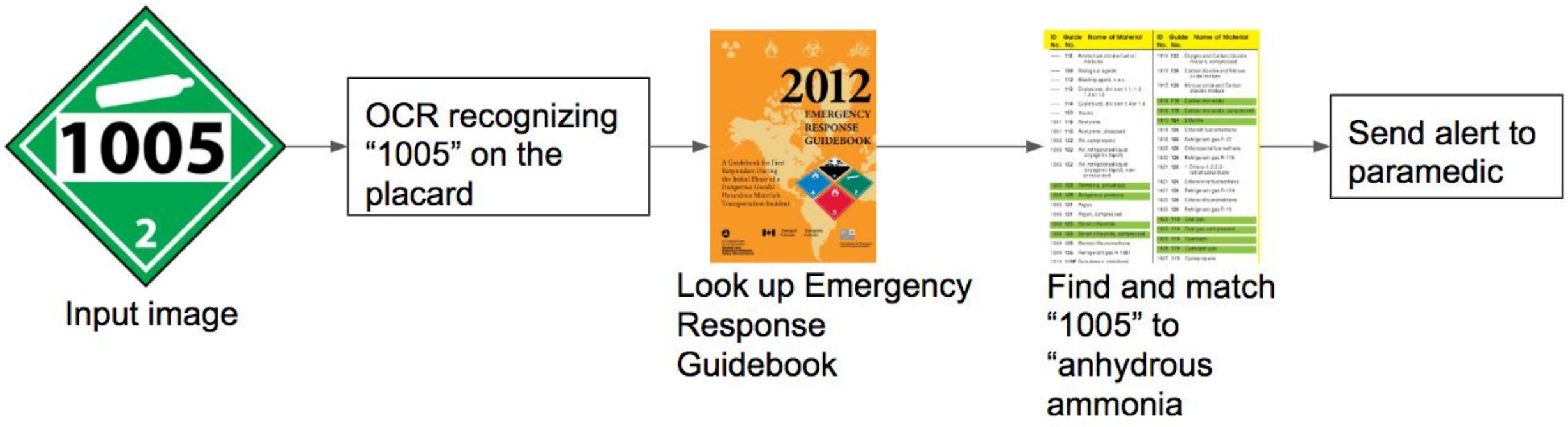

Figure 6. Hazard placard recognition using OCR and matching information from Emergency Response Guidebook (ERG)

## 7. CONCLUSION

In this study, natural language understanding and optical character recognition technology were developed to provide paramedics with the most important situational awareness at an appropriate timing. Specific use cases, including hazard placard recognition, medicine bottle recognition, medication dosage notifications, automatic medical documentation generation (patch form), and automatic first treatment recommendations (standing orders), were especially helpful for paramedics to focus on patient care and reduce errors. However, the technology provided has not been optimized yet, with room for improvement, especially in the user interface and real-time processing capabilities. These design and engineering issues may be addressed in the future commercialization of technology.

## ACKNOWLEDGMENT

The research was carried out at the Jet Propulsion Laboratory, California Institute of Technology, under a contract with the National Aeronautics and Space Administration. The research was funded by the U.S. Department of Homeland Security Science and Technology Directorate Next Generation First Responders Apex Program (DHS S&T NGFR) under NASA prime contract NAS7-03001, Task Plan Number 82-106095.